\documentclass[12pt]{article}
\def\lsim{\mathrel{\lower2.5pt\vbox{\lineskip=0pt\baselineskip=0pt
          \hbox{$<$}\hbox{$\sim$}}}}
\def\gsim{\mathrel{\lower2.5pt\vbox{\lineskip=0pt\baselineskip=0pt
          \hbox{$>$}\hbox{$\sim$}}}}
\begin{document}   
%------------------------------------------------------------------------------
\markright{Living on the edge: cosmology on the boundary of anti-de Sitter space.}
%------------------------------------------------------------------------------
\def\Barcelo{Barcel\'o}
%------------------------------------------------------------------------------
\title{{\bf  Living on the edge: \\
\bf cosmology on the boundary \\
\bf of anti-de~Sitter space}}
\author{Carlos \Barcelo\ and Matt Visser\\[2mm]
{\small \it Physics Department, Washington University, 
Saint Louis, Missouri 63130-4899, USA.}}
\date{{\small 7 April 2000; Revised 18 April 2000; \LaTeX-ed \today}}
%------------------------------------------------------------------------------
\maketitle
%------------------------------------------------------------------------------
\begin{abstract}
We sketch a particularly simple and compelling version of D-brane
cosmology. Inspired by the semi-phenomenological Randall--Sundrum
models, and their cosmological generalizations, we develop a variant
that contains a single (3+1)-dimensional D-brane which is located on
the boundary of a single bulk (4+1)-dimensional region. The D-brane
boundary is itself to be interpreted as our visible universe, with
ordinary matter (planets, stars, galaxies) being trapped on this
D-brane by string theory effects. The (4+1)-dimensional bulk is, in
its simplest implementation, $adS_{4+1}$, anti-de~Sitter space.  We
demonstrate that a $k=+1$ closed FLRW universe is the most natural
option, though the scale factor could quite easily be so large as to
make it operationally indistinguishable from a $k=0$ spatially flat
universe. (With minor loss of elegance, spatially flat and hyperbolic
FLRW cosmologies can also be accommodated.) We demonstrate how this
model can be made consistent with standard cosmology, and suggest some
possible observational tests.

\vspace*{5mm}
\end{abstract}
%-----------------------------------------------------------------------
\vfill
%----------------------------------------------------------------------
\hrule
%-----------------------------------------------------------------------
\bigskip
%-----------------------------------------------------------------------
\centerline{\underline{E-mail:} {\sf carlos@hbar.wustl.edu}}
%-----------------------------------------------------------------------
\centerline{\underline{E-mail:} {\sf visser@kiwi.wustl.edu}}
%-----------------------------------------------------------------------
\bigskip
%-----------------------------------------------------------------------
\centerline{\underline{Homepage:} {\sf http://www.physics.wustl.edu/\~{}carlos}}
%-----------------------------------------------------------------------
\centerline{\underline{Homepage:} {\sf http://www.physics.wustl.edu/\~{}visser}}
%-----------------------------------------------------------------------
\bigskip
%-----------------------------------------------------------------------
\centerline{\underline{Archive:}
{\sf hep-th/0004056}}
%----------------------------------------------------------------------
\bigskip
\hrule
\clearpage
%---------------------------------------------------------------------
% User definitions
%--------------------------------------------------------------------
\def\Box{\nabla^2}
%---------------------------
\def\d{{\mathrm d}}
%----------------------------
\def\ie{{\em i.e.\/}}
\def\eg{{\em e.g.\/}}
\def\etc{{\em etc.\/}}
\def\etal{{\em et al.\/}}
%----------------------------
\def\S{{\mathcal S}}
\def\I{{\mathcal I}}
\def\L{{\mathcal L}}
\def\R{{\mathcal R}}
\def\M{{\mathcal M}}
\def\H{{\mathcal H}}
%-----------------------------
\def\tr{{\mathrm{tr}}}
\def\implies{\Rightarrow}
\def\half{{1\over2}}
%------------------------------
\def\normal{{\mathrm{normal}}}
\def\induced{{\mathrm{induced}}}
\def\stringy{{\mathrm{stringy}}}
\def\graviton{{\mathrm{graviton}}}
\def\critical{{\mathrm{critical}}}
\def\eff{{\mathrm{eff}}}
\def\surface{{\mathrm{surface}}}
\def\bulk{{\mathrm{bulk}}}
\def\boundary{{\mathrm{boundary}}}
\def\matter{{\mathrm{matter}}}
%---------------------------------
\def\Newton{{\mathrm{Newton}}}
\def\Planck{{\mathrm{Planck}}}
\def\Nordstrom{{Nordstr\"om}}
\def\RNdS{Reissner--\Nordstrom--de~Sitter}
%------------------------------------
\def\t{{\hat{t}}}
\def\th{{\hat{\theta}}}
%----------------------------------------------------------------------- 
\def\SIZE{1.00} % for graphics package
%-----------------------------------------------------------------------

%-----------------------------------------------------------------------
\clearpage
%-----------------------------------------------------------------------
\section{Introduction}
%\setcounter{equation}{0}
%------------------------------------------------------------------------------
\label{S:intro}
%------------------------------------------------------------------------------

In this article we develop what we feel is a particularly simple and
compelling cosmological model based on the semi-phenomenological
Randall--Sundrum models for low-energy string theory~\cite{RS1,RS2}.
For some early tentative steps along these lines see the papers of
Gogberashvili~\cite{Gogberashvili1}, plus more recent developments
in~\cite{Gogberashvili2} and~\cite{RS-cosmology}.\footnote{
%--------------------------------------------------------------
Note that many aspects of this recent work can be viewed as extending
domain-wall physics in (3+1) dimensions to brane physics in (4+1)
dimensions, and so owes much to early papers on domain-wall
physics~\cite{Domain-wall}.}
%----------------------------------------------------------------
In developing our cosmology, we wish to minimize the number of baroque
features coming from the underlying string theory, and maximize the
use of symmetry principles, in order to develop a picture that is as
simple and attractive as possible, with good prospects for being
observationally testable.

Perhaps the most compelling model along these lines can be built by
considering a (4+1)-dimensional manifold with a single
(3+1)-dimensional boundary. This boundary is taken to be a D-brane (a
membrane on which the fundamental string fields satisfy Dirichlet type
boundary conditions), and the D-brane is assigned an intrinsic energy
density and pressure arising both from some underlying brane tension
{\em and\/} from matter [ordinary (3+1)-dimensional matter] that is
trapped on the D-brane by stringy effects.

Since this point has the capacity to cause serious confusion, let us
try to make it a little more explicit:\footnote{
%----------------------------------------------------------------
We are trying to make this article comprehensible to string theorists,
relativists, and astrophysicists. Accordingly {\em some} comments may
be trivial to one of the three communities, but we would rather err on
the side of clarity and simplicity than either impenetrable brevity or
excessive technical detail.}
%----------------------------------------------------------------
We are viewing ordinary matter as open-string excitations of the
D-brane boundary. But since open strings by definition have their
end-points on the D-brane, an open string of energy $E$ is strictly
limited in how far it can stretch off the D-brane: Its maximum
extension into any higher-dimensional bulk is simply
$L_\stringy<E/(2\alpha')$ where $\alpha'$ is the fundamental open
string tension.\footnote{
%-------------------------------------------------------------
This whole D-brane picture only makes sense for string excitations of
low energy compared to the string scale: $E<\sqrt{\hbar c \;
\alpha'}$.  So the thickness of the cloud of excitations surrounding
the D-brane is at most of order $L_\stringy<\sqrt{\hbar c
/(2\alpha')}$.}
%-------------------------------------------------------------
In contrast, gravitons are represented by closed string loops which
are not trapped on the D-brane --- gravitons (and non-perturbative
gravity) can very easily penetrate finite distances into the
higher-dimension bulk. Thus gravity is in our model fundamentally a
(4+1)-dimensional effect and we will be using the (4+1)-dimensional
Einstein equations to deduce the analog of the Friedmann equations of
motion for the (3+1)-dimensional D-brane boundary.

Now while gravitons can easily penetrate into the bulk, one does not
want them to be too effective at doing so. Once one turns away from
the large-scale average properties of the cosmological FLRW geometry,
to consider the gravitational field generated by astrophysical
perturbations (planets, stars, galaxies) one does not want the
virtual-graviton cloud surrounding these objects to be completely free
to move into the (4+1)-dimensional bulk, since then one would see an
inverse-cube law for gravity in lieu of the observed inverse-square
law. This is where the Randall--Sundrum mechanism is critical ---
virtual gravitons generated by matter perturbations are (weakly)
trapped near the D-brane, not by stringy effects, but rather by the
bulk gravitational field and the tightly constrained location of the
sources.\footnote{
%-------------------------------------------------------------
The distance scale on which gravitons are trapped is generically set
by the Riemann curvature of the higher-dimensional bulk; in the
Randall--Sundrum models the relevant parameter is $L_\graviton =
\sqrt{6/|\Lambda_{4+1}|}$, defined by the cosmological constant in the
higher-dimensional bulk.}
%-------------------------------------------------------------
We belabor this point because we have seen it generate considerable
confusion within the relativity and astrophysics communities: Gravity
is {\em not\/} used to trap matter on the D-brane and the
Randall--Sundrum models have more in common with the
field-theory-based trapping mechanisms of Akama~\cite{Akama} and
Shaposhnikov~\cite{Shaposhnikov} than they do with the gravity-based
trapping mechanism of~\cite{Exotic}.

In the interests of simplicity and clarity the (4+1)-dimensional bulk
will always be taken to be static and hyper-spherically symmetric,
though we shall quickly specialize to {\RNdS} space, or even more
particularly, to anti-de~Sitter space. The boundary will always be
taken to be hyper-spherically symmetric in the (4+1)-dimensional
sense, with this hyper-spherical symmetry reducing to translation
invariance when viewed from the (3+1)-dimensional point of view. In
picking this particular starting point we have been guided by many
recent publications; including the Randall--Sundrum
scenarios~\cite{RS1,RS2} (which will used to describe the physics near
the brane), the single-brane models of
Gogberashvili~\cite{Gogberashvili1,Gogberashvili2}, various previous
versions of Randall--Sundrum based cosmology~\cite{RS-cosmology}, and
by a desire to have a framework that is at least plausibly connectable
to the complex of ideas going under the name of the adS/CFT
correspondence~\cite{adS/CFT,RS-adS/CFT}.

We start the analysis by a discussion of what it means to apply the
Einstein equations to a {\em manifold with boundary}, interpreting
this process in terms of an extension of the Israel--Lanczos--Sen thin
shell formalism~\cite{Israel,Lanczos,Sen}. This permits us to write
down an analog of the usual Friedmann equation of FLRW cosmology, and
in the next section we discuss how to make this cosmologically
viable. Going beyond the FLRW cosmological fluid approximation we
verify that the essential portion of the Randall--Sundrum model
(having to do with the weak trapping of perturbatively generated
gravitons near the brane) continues to work in the present
context. Finally we indicate some possible variants on the present
model and describe areas where the present ideas may lead to
observational tests.

%-----------------------------------------------------------------------------
\section{D-brane surgery}
%\setcounter{equation}{0}
%------------------------------------------------------------------------------
\label{S:surgery}
%------------------------------------------------------------------------------
\subsection{Extrinsic and intrinsic geometries:}
%------------------------------------------------------------------------------

We start by considering a rather general static hyper-spherically
symmetric geometry in (4+1) dimensions. (This is not the most general
such metric, but quite sufficient for our purposes.)\footnote{
%----------------------------------------------------------------
Note that the technical computations closely parallel those for
spherically symmetric (2+1)-dimensional domain walls symmetrically
embedded in a spherically symmetric (3+1)-dimensional spacetime. See,
for instance, references~\cite{Book,Surgery,Brane-surgery}.}
%-----------------------------------------------------------------
%
\begin{equation}
\label{E:bulk}
\d s^2_{4+1} = - F(r) \; \d t^2 +{\d r^2\over F(r)} + r^2 \; \d\Omega_3^2.
\end{equation}
\begin{equation}
\d\Omega_3^2 \equiv \d\chi^2 + 
\sin^2\chi \; (\d\theta^2 + \sin^2\theta \; \d\phi^2).
\end{equation}
To build the class of (3+1)-dimensional geometries we are interested
in, we start by simply truncating the (4+1)-dimensional geometry at
some time-dependent radius $a(t)$, keeping only the interior portion
and discarding the exterior.  Kinematically, the surface of this
truncated geometry (which we take to be the location of the D-brane)
is automatically a (3+1)-dimensional closed ($k=+1$, positive spatial
curvature) FLRW geometry with induced metric
\begin{equation}
\d s^2_{3+1} = 
-
\left[ F(a(t)) - {1\over F(a(t))} \left({\d a\over\d t}\right)^2 \right]\; \d t^2 
+ a(t)^2  \; \d\Omega_3^2.
\end{equation}
Now consider radial motion of the D-brane; this is radial motion in
the embedding (4+1)-dimensional hyperspace. We start the analysis by
first parameterizing the motion in terms of proper time along a curve
of fixed $\chi$, $\theta$, and $\phi$ (these are comoving coordinates
in the FLRW cosmology). That is: the D-brane sweeps out a world-volume
\begin{equation}
X^\mu(\tau,\chi,\theta,\phi) = \left(t(\tau),a(\tau),\chi,\theta,\phi\right).
\end{equation}
The 5-velocity of the $(\chi,\theta,\phi)$ element of the D-brane can then
be defined as
\begin{equation}
V^\mu = \left({\d t \over\d \tau}, {\d a\over\d\tau},0,0,0\right).
\end{equation}
Using the normalization condition and the assumed form of the metric,
and defining $\dot a = \d a/\d\tau$,
\begin{equation}
V^\mu = \left({\sqrt{F(a)+\dot a^2}\over F(a)},\dot a, 0,0,0\right); \qquad 
V_\mu =
\left(-\sqrt{F(a)+\dot a^2},{\dot a\over F(a)},0,0,0\right).
\end{equation}
The unit normal vector to the hypersphere $a(\tau)$ is
\begin{equation}
n^\mu = \left(-{\dot a\over F(a)},-\sqrt{F(a)+ \dot a^2},0,0,0\right); \qquad 
n_\mu = \left(+\dot a,-{\sqrt{F(a)+\dot a^2}\over F(a)},0,0,0\right).
\end{equation}
[We shall take the unit normal to be inward pointing, into the bulk of
the (5+1) geometry.]  The extrinsic curvature can be written in terms
of the normal derivative\footnote{
%-------------------------------------------------
Unfortunately sign conventions differ on this point. 
We follow~\cite{Book}.}
%-------------------------------------------------
%
\begin{equation}
K_{\mu\nu} 
= \half  {\partial g_{\mu\nu}\over\partial \eta}
= \half \; n^\sigma \;  {\partial g_{\mu\nu}\over\partial x^\sigma}.
\end{equation}
If we go to an orthonormal basis, the $\hat{\chi}\hat{\chi}$ component
is easily evaluated
\begin{equation}
K_{\hat{\chi}\hat{\chi}} = K_{\th\th} = K_{\hat{\phi}\hat{\phi}} 
= - \half  \sqrt{F(a)+\dot a^2} \; {\partial g_{\chi\chi}\over\partial r} \; 
   g^{\chi\chi} 
= - {\sqrt{F(a)+\dot a^2}\over a}
\end{equation}
The $\tau\tau$ component is a little messier, but generalizing the
calculation of~\cite{Surgery} (which amounts to calculating the
five-acceleration of the brane, this is explained in more detail
in~\cite{Book}) quickly leads to
\begin{equation}
K_{\hat{\tau}\hat{\tau}} 
= + \half  {1\over\sqrt{F(a)+\dot a^2}} \; 
\left({\d F(r)\over\d a} + 2 \ddot a \right)
= + {\d\over\d a} \left(\sqrt{F(a)+\dot a^2}\right) .
\end{equation}
In contrast to the extrinsic geometry, the intrinsic geometry of the
D-brane is in these coordinates simply
\begin{equation}
\d s^2_{3+1} = - \d\tau^2 + a(\tau)^2  \; \d\Omega_3^2.
\end{equation}
%

%--------------------------------------------------
\subsection{The D-brane as boundary:}
%--------------------------------------------------

A perhaps unusual (and for us very useful) feature of D-brane physics
is that the D-brane can be viewed as an actual physical boundary to
spacetime, with the ``other side'' of the D-brane being empty (null and
void).\footnote{
%------------------------------------------------------------------
A brief sketch of these ideas, from the (3+1)-dimensional point of
view where one is dealing with holes in spacetime (voids), was
presented in~\cite{Brane-surgery}. Here we expand on these ideas in a
more explicit manner.}
%------------------------------------------------------------------
In general relativity, as it is normally formulated, the notion of an
actual physical boundary to spacetime (that is, an accessible boundary
reachable at finite distance) is complete anathema. The reason that
spacetime boundaries are so thoroughly deprecated in general
relativity is that they are artificial special places in the manifold
where some sort of boundary condition has to be placed on the
physics. Without such a postulated boundary condition all
predictability is lost, and the theory is not physically
acceptable. Since without some deeper underlying theory there is no
physically justifiable reason for picking any one particular type of
boundary condition (Dirichlet, Neumann, Robin, or something more
complicated), the attitude in standard general relativity has been to
simply exclude boundaries.

The key difference when a D-brane is used as a boundary is that now
there is a specific and well-defined boundary condition for the
physics: D-branes (remember that ``{\em D is for Dirichlet\/}'') are
defined as the loci on which the fundamental open strings end (and
satisfy Dirichlet-type boundary conditions). D-branes are therefore
capable of providing both a physical boundary for the spacetime {\em
and} a plausible boundary condition for the physics residing in the
spacetime.\footnote{
%----------------------------------------------------------------------
A word of warning: D-branes by definition provide boundary conditions
directly on the fundamental string states, and so, since all physics
in string theory can be viewed in terms of some combination of string
states, D-branes will {\em in principle} provide boundary conditions
for all the physics. In practice the route from string state to
low-energy effective field theory may be rather indirect, and
elucidation of the proper boundary condition may be a little obscure;
when in doubt use symmetry as much as possible, and be prepared to
keep at least a few adjustable constants as part of the low-energy
semi-phenomenological theory.}
%---------------------------------------------------------------------

When it comes to specific calculations, this is however not be the
best mental picture to have in mind --- after all, how would you try
to calculate the Riemann tensor for the edge of spacetime? And what
would happen to the Einstein equations at the edge? There is a
specific technical trick that clarifies the situation: Take the
manifold with D-brane boundary and make a second copy (including a
second D-brane boundary), then sew the two manifolds together along
their respective D-brane boundaries, creating a single manifold
without boundary that contains the doubled D-brane, and exhibits a
$Z_2$ symmetry on reflection around the D-brane. Because this new
manifold is a perfectly reasonable no-boundary manifold containing a
(thin shell) D-brane, the gravitational field can be analyzed using a
slight generalization of the usual Israel--Lanczos--Sen thin-shell
formalism of general relativity~\cite{Israel,Lanczos,Sen}. We now need
to consider (3+1) shells propagating in (4+1) space, but this merely
changes a few integer coefficients. The metric is continuous, the
connection exhibits a step-function discontinuity, and the Riemann
curvature a delta-function at the D-brane. The dynamics of the D-brane
can then be investigated in this $Z_2$-doubled manifold, and once the
dynamical equations and their solutions have been investigated the
second surplus copy of spacetime can quietly be forgotten (effectively
halving the strength of the delta-function contribution to the Riemann
tensor). That is, as long as one is working in the $Z_2$-doubled
manifold the discontinuity in the extrinsic curvature is twice the
extrinsic curvature as seen from either side
\begin{equation}
\kappa_{\alpha\beta}(Z_2) = [K_{\alpha\beta}] = 
K^+_{\alpha\beta} - K^-_{\alpha\beta} = 2 K_{\alpha\beta}.
\end{equation}
Consequently the Riemann tensor in the $Z_2$-doubled manifold is
\begin{eqnarray}
R(Z_2)_{\alpha\beta\gamma\delta} &=& 
- 2 \delta(\eta) \; \left[
K_{\alpha\gamma} \; n_\beta \; n_\delta + 
K_{\beta\delta} \; n_\alpha \; n_\gamma -
K_{\alpha\delta} \; n_\beta \; n_\gamma -
K_{\beta\gamma} \; n_\alpha \; n_\delta
\right] 
\nonumber\\
&&
+\Theta(\eta) \; R^+_{\alpha\beta\gamma\delta}
+\Theta(-\eta) \; R^-_{\alpha\beta\gamma\delta}.
\end{eqnarray}
We now {\em define} the Riemann tensor of the manifold with boundary
by throwing away half of the $Z_2$-doubled manifold, and in view of
the manifest symmetry of the situation, also throwing away half the
delta-function contribution.\footnote{
%-----------------------------------------------------------------
If for whatever reason one does not wish to work with the
$Z_2$-doubled manifold, there is an alternative construction that
leads to the same result that we present in Appendix A.}
%----------------------------------------------------------------- 

After doing all this, near the D-brane boundary the Riemann
tensor takes the form
\begin{equation}
R_{\alpha\beta\gamma\delta} = 
- \delta(\eta) \; \left[
K_{\alpha\gamma} \; n_\beta \; n_\delta + 
K_{\beta\delta} \; n_\alpha \; n_\gamma -
K_{\alpha\delta} \; n_\beta \; n_\gamma -
K_{\beta\gamma} \; n_\alpha \; n_\delta
\right] +
R^\bulk_{\alpha\beta\gamma\delta}.
\end{equation}
This is the relevant generalization of equation (14.23) of~\cite{Book}
to a manifold with boundary; note that there is only one side to the
boundary and that we explicitly use only the extrinsic curvature of
that one side (which is half the extrinsic curvature discontinuity in
the $Z_2$-doubled manifold). This particular formula is valid for any
([n--1]+1)-dimensional boundary to a (n+1)-dimensional bulk. It does
assume that the normal $n$ is spacelike, though no symmetry
assumptions are made. If we introduce the general projection tensor
\begin{equation}
h_{\mu\nu} = g_{\mu\nu} - n_\mu \; n_\nu,
\end{equation}
then this projection tensor {\em is} the induced metric on the
boundary and in the particular application we have in mind will be the
physical spacetime metric of our universe.  Performing the relevant
contractions, and still working in an arbitrary number of bulk
dimensions
\begin{equation}
R_{\mu\nu} = - \delta(\eta) \; 
\left[ K_{\mu\nu} + K \; n_\mu \; n_\nu \right] + 
R^\bulk_{\mu\nu};
\end{equation}
\begin{equation}
R = -2 K \; \delta(\eta) + R^\bulk;
\end{equation}
\begin{equation}
G_{\mu\nu} = - \delta(\eta) \;
\left[ K_{\mu\nu} - K \; h_{\mu\nu} \right] + G^\bulk_{\mu\nu}.
\end{equation}
These formulae generalize (14.25)--(14.27) of~\cite{Book} to a
manifold with boundary. With hindsight this makes perfectly good sense
since if we now integrate over the complete manifold (bulk plus
boundary)
\begin{eqnarray}
\oint \d^{n+1} x \; \sqrt{-g_{n+1}} \; R 
&=& 
\int_\bulk \d^{n+1} x \; \sqrt{-g_{n+1}} \; R_\bulk 
\nonumber\\
&-&
2 \int_\boundary \d^{[n-1]+1} x \; \sqrt{-g_{[n-1]+1}} \; K.
\end{eqnarray}
Which means that we have automatically recovered the Gibbons--Hawking
surface term for the gravitational action, in addition to the
Einstein--Hilbert bulk term.

We also take the total stress-energy tensor to be given by a
combination of surface and bulk components
\begin{equation}
T_{\mu\nu} = \delta(\eta) \;  T^\surface_{\mu\nu} + T^\bulk_{\mu\nu},
\end{equation}
and normalize our (n+1)-dimensional bulk Newton constant $G_{n+1}$ by
\begin{equation}
G_{\mu\nu} = 8\pi G_{n+1} \; T_{\mu\nu}.
\end{equation}
Then in particular, picking off the surface contribution to both the
Einstein tensor and the stress-energy
\begin{equation}
8 \pi G_{n+1} \; T^\surface_{\mu\nu} 
= - \left[ K_{\mu\nu} - K \; h_{\mu\nu} \right]
\end{equation}
Whether or not this surface stress tensor satisfies the energy
conditions depends on the signs of the eigenvalues of the extrinsic
curvature. By looking at the $Z_2$-doubled geometry it is a general
result~\cite{Book} that a convex boundary (when viewed from the bulk)
violates the null energy condition (NEC), while a concave boundary
satisfies it. (This is intimately related to the fact that traversable
wormholes violate the null energy condition,
see~\cite{Book,Surgery,Morris-Thorne,Hochberg,Survey,Examples}.)

%---------------------------------------------
%\clearpage
%---------------------------------------------
\subsection{Cosmology}
%---------------------------------------------

Now particularize to the (4+1)-dimensional version of the thin-shell
formalism, and use the FLRW symmetries of the D-brane (some of the
integer coefficients and exponents appearing below are dimension
dependent):
\begin{equation}
8\pi \; G_{4+1} \; \rho_{3+1} 
=  3  \; {\sqrt{F(a)+\dot a^2}\over a}.
\end{equation}
(Note that the energy density is positive definite, in agreement with
the fact that this boundary is concave when viewed from the
bulk.)\footnote{
%--------------------------------------------------------------
Because of this feature the D-brane occurring here is guaranteed to
have positive tension, and we do not need to worry (at least not at
the cosmological level) about the possibility of
energy-condition-violating negative tension D-branes, and the somewhat
peculiar features [traversable wormholes, \etc] that negative tension
D-branes can introduce into the low energy effective
theory~\cite{Brane-surgery}.}
%--------------------------------------------------------------
%
\begin{equation}
8\pi \; G_{4+1} \; p_{3+1}
=   - {1\over a^2}  
\; {\d\over\d a} \left(a^2 \sqrt{F(a)+\dot a^2} \right).
\end{equation}
These equations can easily be seen to be compatible with the
conservation of the stress energy localized on the D-brane\footnote{
%--------------------------------------------------------------------
There is another potential source of confusion here: Since the
(3+1)-dimensional D-brane is sweeping through the (4+1)-dimensional
bulk, why is it that the D-brane does not exchange energy with the
bulk?  One might at first glance expect violations of
(3+1)-dimensional stress-energy conservation due to (4+1)-dimensional
matter entering or leaving the D-brane. In fact, in general this might
happen, and it is potentially an interesting observational signal to
look for --- but in the present cosmological context this effect is
zero: as the D-brane moves through (4+1)-space, it is the ``flux'' of
(4+1)-dimensional matter onto the brane, defined by
\[
J_\mu = 
n_\alpha \; T^{\alpha\beta} \; \left[ g_{\beta \mu} - n_\beta \; n_\mu  \right],
\]
that determines whether or not (4+1)-dimensional stress-energy
conservation holds~\cite{Book}. In all of the bulk geometries
considered in this article, this flux is identically zero (in fact the
stress-energy tensor is diagonal).}
%--------------------------------------------------------------------
%
\begin{equation}
{\d\over\d\tau} (\rho_{3+1} \; a^3) + p_{3+1} \; {\d\over\d\tau} (a^3) = 0.
\end{equation}
So as usual, {\em two} of these three equations are independent, 
and the third is redundant.

The conservation equation is identical to that for standard cosmology,
while the D-brane version of the Friedmann equation, obtained by
rearranging the equation for the surface energy density that was given
above, is seen to be
\begin{equation}
\left({\dot a \over a}\right)^2 = 
- {F(a)\over a^2} + \left({ 8\pi \; G_{4+1} \; \rho_{3+1}\over 3}\right)^2.
\end{equation}
In contrast the {\em standard} Friedmann equation (for a $k=+1$ closed
FLRW universe) is
\begin{equation}
\left({\dot a \over a}\right)^2 = 
- {1\over a^2} + {\Lambda\over3}+ { 8\pi \; G_{3+1} \; \rho\over 3}.
\end{equation}
To get a brane cosmology that is not wildly in conflict with
observation, we split the (3+1)-dimensional energy into a constant
$\rho_0$ determined by the brane tension, plus ordinary matter $\rho$,
with $\rho\ll\rho_0$ to suppress the quadratic term in comparison to
the linear~\cite{RS-cosmology}. Then with $\rho_{3+1} = \rho_0 + \rho$
we have
\begin{equation}
\left({\dot a \over a}\right)^2 = 
- {F(a)\over a^2} + 
\left({ 8\pi \; G_{4+1} \; \rho_0\over3}\right)^2  + 
\left({ 16\pi \; G_{4+1} \; \rho_0\over 3}\right)
\left({ 8\pi \; G_{4+1} \;\over 3}\right)  
\left[\rho + \half {\rho^2\over\rho_0}\right].
\end{equation}
Picking out the term linear in $\rho$, this permits us to identify
\begin{equation}
G_{3+1} = G_{4+1} \left({ 16\pi \; G_{4+1} \; \rho_0\over 3}\right); 
\qquad \hbox{that is} \qquad
G_{4+1} = \sqrt{3\;G_{3+1}\over16\pi\; \rho_0}.
\end{equation}
Therefore
\begin{equation}
\label{E:formal}
\left({\dot a \over a}\right)^2 = 
- {F(a)\over a^2} + 
\left({ 8\pi \; G_{3+1}\over3}\right) 
\left[\half \rho_0 + \rho + \half {\rho^2\over\rho_0} \right]. 
\end{equation}
Since we want $\rho_0\gg\rho$ to suppress the quadratic term, this
leaves us with a large (3+1)-dimensional cosmological constant that we
will need to eliminate by cancelling it (either fully or partially)
with some term in $F(a)$~\cite{RS-cosmology}.  This result is in its
own way quite remarkable: up to this point no assumptions had been
made about the size of the brane tension, or even whether or not the
brane tension was zero. Nor had any assumption been made up to this
point about the existence or otherwise of any cosmological constant
in the (4+1)-dimensional bulk. It is observational cosmology that
first forces us to take $\rho_0$ large [electro-weak scale or higher
to avoid major problems with nucleosynthesis], and then forces us to
deduce the presence of an almost perfectly countervailing cosmological
constant in the bulk~\cite{RS-cosmology}.

In the next section we shall make use of this still relatively general
formalism by specializing $F(r)$ to the {\RNdS} form.

%------------------------------------------------------------------------------
\subsection{Reissner--\Nordstrom--de~Sitter surgery}
%------------------------------------------------------------------------------
\label{SS:RNdS}
%------------------------------------------------------------------------------

For the (4+1)-dimensional {\RNdS} geometry
\begin{equation}
F(r) = 1 - {2M_{4+1}\over r^2} 
+ {Q_{4+1}^2\over r^4} - {\Lambda_{4+1} \; r^2\over 6}.
\end{equation}
Here $M_{4+1}$ is a (4+1)-dimensional ``mass'' parameter,
corresponding to the mass of the central object in (4+1)-space --- it
does not have a ready (3+1)-dimensional interpretation and is best
carried along as an extra free parameter that from the 4-dimensional
point of view can be adjusted to taste. Similarly, $Q_{4+1}$
corresponds to an ``electric charge'' in the (4+1)-dimensional
sense. Our universe, the boundary D-brane, must then be viewed as
carrying an equal but opposite charge to allow field lines to
terminate. From the (3+1)-dimensional view this may be taken to be a
second free parameter. The (4+1)-dimensional cosmological constant
combines with the term coming from the D-brane tension to give an
effective (3+1)-dimensional cosmological constant
\begin{equation}
\Lambda = {\Lambda_{4+1}\over2} + 4\pi G_{3+1}\; \rho_0.
\end{equation}
In the original Randall--Sundrum models these two terms were
fine-tuned by hand to obtain complete cancellation. In view of the
recent observational evidence for a small cosmological constant in our
observable universe we need merely assert that this effective
cosmological constant is presently relatively small ($\Lambda \lsim
8\pi G_{3+1}\; \rho_\critical$; this is small by particle physics
standards, but can be quite significant by cosmological
standards).\footnote{
%-------------------------------------------------------------
A small effective cosmological constant would indeed imply deviations
from the original Randall--Sundrum scenario, but on a distance scale
determined by this effective cosmological constant (and observationally
this distance scale would be of order Giga-parsecs or larger). So we
are not too concerned about this issue in that the implications for
particle phenomenology are negligible.}
%-------------------------------------------------------------

Since $\rho_0$ is guaranteed positive,\footnote{
%-------------------------------------------------------------
Tricky point: actually it is $\rho_{3+1}$ that is guaranteed to be
positive, and this holds because the D-brane universe is taken to be
convex as seen from the bulk. Then the same logic that leads to energy
condition violations for traversable wormholes now applies in reverse,
and the (3+1) null energy condition is generically satisfied in this
type of cosmological model. (Violating the strong energy condition,
which is relevant for cosmological inflation, is much
easier~\cite{Science}.)}
%-------------------------------------------------------------
this implies that $\Lambda_{4+1}$ should be negative, and so if this
model is correct we are living on the edge of a bulk anti-de~Sitter
space. The D-brane dynamical equation now reads
\begin{equation}
\left({\dot a \over a}\right)^2 = 
- {1\over a^2} 
+ {2M_{4+1}\over a^4} 
-  {Q_{4+1}^2\over a^6} 
+ {\Lambda\over 3}\;
+\left({ 8\pi \; G_{3+1}\over3}\right) 
\left[\rho + \half {\rho^2\over\rho_0} \right]. 
\end{equation}
It is clear that by tuning these parameters appropriately one can
recover standard cosmology to arbitrary accuracy. The $M_{4+1}$
parameter can be used to mimic an arbitrary quantity of what would
usually be called ``radiation'' (relativistic fluid, $\rho=3p$), while
the $Q_{4+1}$ parameter mimics ``stiff'' matter
($\rho=p$)~\cite{Science}, though with an overall minus sign. An
observational astrophysicist or cosmologist could now simply forget
about the underlying string theory and D-brane physics, take this
expression as the D-brane inspired generalization of the Friedmann
equations, and treat $M_{4+1}$, $Q_{4+1}$, $\Lambda$, and $\rho_0$ as
parameters to be observationally determined.

Since we actually want to do more than just reproduce standard
cosmology we should seek some additional constraints on these
parameters --- and this is where the phenomenon of weak localization
of the graviton near the brane comes into play.

%------------------------------------------------------------------------------
\section{Weak localization of perturbative gravity}
%\setcounter{equation}{0}
%------------------------------------------------------------------------------
\label{S:weak}
%------------------------------------------------------------------------------

Suppose that the observable universe is large compared to the natural
distance scale in the (4+1)-dimensional bulk, that is: $a(\tau)
\gg\sqrt{6/|\Lambda_{4+1}|}$ (so that the universe has ``grown up''),
and both $M_{4+1}$ and $Q_{4+1}$ are sufficiently small to allow us to
approximate
\begin{equation}
F(r) \approx { |\Lambda_{4+1}| \; r^2 \over 6}; 
\qquad \hbox{for} \qquad 
r \approx a.
\end{equation}
Then near the D-brane we can write
\begin{equation}
\d s^2_{4+1} \approx - { |\Lambda_{4+1}| \; r^2 \over 6} \; \d t^2 
+{6\over |\Lambda_{4+1}| \; r^2} \; \d r^2 + r^2 \; \d\Omega_3^2.
\end{equation}
In terms of the normal distance (proper distance) from the D-brane,
\begin{equation}
\eta \approx \sqrt{6\over |\Lambda_{4+1}|} \; \ln(r/a),
\end{equation}
this implies
\begin{equation}
\d s^2_{4+1} \approx 
+\d\eta^2 
+ \exp\left(-2\sqrt{|\Lambda_{4+1}|\over6}\; \eta\right)
\left[ - { |\Lambda_{4+1}| \; a^2\over 6} \; \d t^2  + a^2 \; \d\Omega_3^2 \right].
\end{equation}
If we now re-label our time parameter in terms of proper time measured
along the D-brane (that is, use the proper time of a cosmologically
comoving observer),
\begin{equation}
\tau \approx \sqrt{|\Lambda_{4+1}| \; a^2 \over6} \; t,
\end{equation}
and introduce quasi-Cartesian coordinates to the tangent space at any
arbitrary point point of the D-brane then
\begin{equation}
\d s^2_{4+1} \approx 
+\d\eta^2 
+ \exp\left(-2\sqrt{|\Lambda_{4+1}|\over6}\; \eta\right)
\left[ - \d\tau^2  + \d x^2 + \d y^2 + dz^2 \right].
\end{equation}
Thus in this approximation the near-brane metric is precisely of the
Randall--Sundrum form~\cite{RS1,RS2} and we know from their analysis
that there is a graviton bound state attached to the brane with an
exponential falloff controlled by the distance scale
parameter\footnote{
%----------------------------------------------------------------
Of course this is little more than the statement that if we are
interested in laboratory physics in the here and now, then a tangent
space approximation to cosmology had better work: Minkowski space is
an excellent approximation for physics here on Earth and so the
D-brane {\em must\/} exhibit at least approximate Lorentz symmetry if
it is to be acceptable as a model of empirical reality. Moving off the
D-brane and into the bulk, the only essential item is that at large
enough distances [from the (4+1) ``centre''] we must have $F(r)\propto
r^2$. Thus as long as the (4+1) geometry is asymptotically
anti-de~Sitter space we will recover Randall--Sundrum phenomenology on
small scales. (And eventually, on large enough distance scales, the
simple Randall--Sundrum phenomenology will break down either because of
cosmological expansion, or because of the small effective
(3+1)-dimensional cosmological constant, or simply because of the
positive spatial curvature [$k=+1$ and $a$ is finite].)  }
%-----------------------------------------------------------------
%
\begin{equation}
L_\graviton = \sqrt{6\over|\Lambda_{4+1}|}.
\end{equation}
Now the experimental fact that we do not see short distance deviations
from the inverse square law of gravity at least down to centimetre
scales implies that $L_\graviton$ is certainly less than one
centimetre (and many would argue that it is at most one
millimetre). Numerous experiments designed to tighten this limit are
currently planned and in progress. Within the approximation that the
(3+1)-dimensional effective cosmological constant is negligible we get
\begin{equation}
G_{3+1} = G_{4+1} \; {2\over L_\graviton}.
\end{equation}
The importance of these results for cosmology is that, given the
observed almost perfect cancellation of the net cosmological constant,
\begin{equation}
\rho_0 \approx {3\over4\pi\;G_{3+1}\;L_\graviton^2} 
= {3\over4\pi} {L_\Planck^2\over L_\graviton^2}\; \rho_\Planck 
\end{equation}
While this number is certainly large on a usual astrophysics scale,
and is rather large even compared with nuclear densities, it could
still be much less than the Planck scale and yet be compatible with
experiment. Indeed if $L_\graviton$ is as large as a centimetre then
the quadratic terms in the density become important once temperatures
reach the electroweak scale (about 100 GeV).  The good news is that
this implies the model is compatible with standard cosmology at least
back to the electroweak scale; the better news is that there are
possibilities of seeing deviations from the standard cosmology as we
go further back. The larger $L_\graviton$ is, the better things are
with regard to the hierarchy problem of particle physics~\cite{RS1,RS2}
and the lower the brane tension needs to be. On the other hand, the
lower $L_\graviton$ is the better the brane is at trapping
gravitational perturbations and the less risk there is of conflict
with gravity-based experiment.

%-----------------------------------------------------------------------------
%\clearpage
%------------------------------------------------------------------------------
\section{Discussion}
%\setcounter{equation}{0}
%------------------------------------------------------------------------------
\label{S:discusion}
%------------------------------------------------------------------------------

The Randall--Sundrum scenarios~\cite{RS1,RS2}, and earlier tentative
steps along these lines~\cite{Gogberashvili1}, have engendered a
tremendous amount of activity, both in terms of particle physics and
in terms of cosmology~\cite{Gogberashvili2,RS-cosmology}. In this
paper we have sketched what we feel is perhaps the simplest most
symmetric cosmology that can be based on these ideas: We have reduced
the number of D-branes to exactly one, and have only one bulk
(4+1)-dimensional region. The D-brane (which our observable universe
lives on) is here viewed as an actual physical boundary to the
higher-dimensional spacetime, and we have demonstrated how to write
down both curvature tensor and field equations for a manifold with
boundary.

We have verified that standard $k=+1$ FLRW cosmology can very easily
be reproduced, and that we do not have massive present day violations
of observational constraints.  If you absolutely insist on a spatially
flat $k=0$ geometry (or even a spatially hyperbolic $k=-1$ geometry)
that can also be achieved along the lines of this article, but at some
cost in elegance, and for very little real purpose. Remember that for
$a(\tau)$ large enough a $k=+1$ spatial slice mimics $k=0$ to
arbitrary accuracy. In Appendix B and Appendix C we sketch how one
could nevertheless force spatially flat or spatially hyperbolic FLRW
cosmologies into this framework.

As is by now not unexpected~\cite{RS1,RS2,RS-cosmology}, likely
places to look for observational signatures are in short-distance
(centimetre) deviations from the gravitational inverse square law, and
in very early universe cosmology (before densities drop to the
electro-weak scale; this is the region where the quadratic density
term in the generalized Friedmann equation might come into
play).\footnote{
%-----------------------------------------------------------
In particular, for $\rho \gg \rho_0$ even ordinary radiation
($\rho\propto a^{-4}$) acts as though it has a $a^{-8}$ behaviour, and
this is enough to drive an epoch of power-law inflation with
$a(t)\propto t^{1/4}$~\cite{RS-cosmology}.}
%-----------------------------------------------------------

Because we are viewing the D-brane as an actual boundary, the
conjectured connections between the Randall--Sundrum models and
Maldacena's adS/CFT conjecture are perhaps more
compelling~\cite{adS/CFT,RS-adS/CFT} --- we no longer have to deal
with a $Z_2$-doubled version of the adS/CFT conjecture, but can work
directly on a boundary of the (asymptotic) anti-de~Sitter space. As
the universe evolves in time the D-brane boundary moves further out
into the asymptotic anti-de~Sitter region, and this hints at a
possible connection between cosmological time, the holographic
hypothesis, and renormalization group flow~\cite{RS-adS/CFT}.

%------------------------------------------------------------------------------
\section*{Acknowledgments}
%------------------------------------------------------------------------------

The research of CB was supported by the Spanish Ministry of Education
and Culture (MEC). MV was supported by the US Department of Energy.

%---------------------------------------------------------------
\appendix
%---------------------------------------------------------------
\section*{Appendix A: Alternative construction for the \\
Riemann tensor of a manifold with boundary.}
%\setcounter{section}{A}
%\setcounter{equation}{0}
%--------------------------------------------------------------

Take your original manifold $\M$, with boundary $\partial\M$, and join
to the boundary a hyper-tube of topology $\H = (-\infty,0) \otimes
\partial\M$. Let the metric on this hyper-tube be specified in terms
of the induced metric on the boundary and the flat 1-dimensional metric:
\begin{equation}
g(\H) = d\eta^2 \oplus g(\partial\M).
\end{equation}
Then by construction $K^-_{\alpha\beta} =0$ and $K^+_{\alpha\beta} =
K_{\alpha\beta}$, so that in this geometry
\begin{equation}
\kappa_{\alpha\beta}(\M\cup\H) = [K_{\alpha\beta}] = 
K^+_{\alpha\beta} - K^-_{\alpha\beta} =  K_{\alpha\beta},
\end{equation}
leading to the Riemann tensor~\cite{Book} 
\begin{eqnarray}
R(\M\cup\H)_{\alpha\beta\gamma\delta} &=& 
- \delta(\eta) \; \left[
K_{\alpha\gamma} \; n_\beta \; n_\delta + 
K_{\beta\delta} \; n_\alpha \; n_\gamma -
K_{\alpha\delta} \; n_\beta \; n_\gamma -
K_{\beta\gamma} \; n_\alpha \; n_\delta
\right] 
\nonumber\\
&&
+\Theta(\eta) \; R(\M)^\bulk_{\alpha\beta\gamma\delta}
+\Theta(-\eta) \; R(\H)^\bulk_{\alpha\beta\gamma\delta}.
\end{eqnarray}
Now truncate the geometry by simply throwing away the hyper-tube $\H$.
The Riemann tensor in the remaining manifold $\M$ is, as before
\begin{equation}
R(\M)_{\alpha\beta\gamma\delta} = 
- \delta(\eta) \; \left[
K_{\alpha\gamma} \; n_\beta \; n_\delta + 
K_{\beta\delta} \; n_\alpha \; n_\gamma -
K_{\alpha\delta} \; n_\beta \; n_\gamma -
K_{\beta\gamma} \; n_\alpha \; n_\delta
\right] +
R(\M)^\bulk_{\alpha\beta\gamma\delta}.
\end{equation}
There is now no symmetry to suggest that one should perform any
particular splitting of the delta-function contribution at the
boundary, and in fact the observation that the extrinsic curvature is
by construction zero on the hyper-tube side of the boundary is an
indication that you should assign all the delta-function contribution
to $\M$, the resulting manifold with boundary. Either construction
(hyper-tube addition or $Z_2$-doubling) leads to the same result for
the Riemann tensor, but some may be happier with one construction over
the other.

%---------------------------------------------------------------
\section*{Appendix B: Spatially flat FLRW cosmology.}
%\setcounter{section}{B}
%\setcounter{equation}{0}
%---------------------------------------------------------------

By a little guess-work based on hyper-spherically symmetric {\RNdS}
space one is led to consider the metric
\begin{equation}
\d s^2_{4+1} = 
-  F(r)\; \d t^2 
+{\d r^2\over F(r)} 
+ r^2 \; \left[\d x^2 + \d y^2 + \d z^2 \right].
\end{equation}
with (note the {\em absence} of the leading $1$!)
\begin{equation}
F(r) =  - {2M_{4+1}\over r^2} 
+ {Q_{4+1}^2\over r^4} - {\Lambda_{4+1} \; r^2\over 6}.
\end{equation}
This metric still satisfies the (4+1)-dimensional Einstein--Maxwell
equations, but with a hyper-planar symmetry instead of a
hyper-spherical symmetry. You can now re-do the analysis of this note
by placing a spatially flat D-brane boundary at $r=a(t)$ and will
obtain very similar results to those of this article. The intrinsic
geometry of the D-brane will now be
\begin{equation}
\d s^2_{3+1} = - \d \tau^2  + a^2(\tau) \; \left[\d x^2 + \d y^2 + \d z^2 \right].
\end{equation}
It is not clear to us that the marginal change in the Friedmann
equation is worth the loss of hyper-spherical symmetry. The point
$r=0$ is still (for $M_{4+1}\neq0$ or $Q_{4+1}\neq0$) a curvature
singularity of the (4+1)-dimensional bulk, but whether you really want
to call it the ``center'' of the bulk (as opposed to say a ``focal
point'') is somewhat less than clear.

%---------------------------------------------------------------
\section*{Appendix C: Spatially hyperbolic FLRW cosmology.}
%\setcounter{section}{C}
%\setcounter{equation}{0}
%--------------------------------------------------------------

Inspired by the previous guess-work one is led to consider the metric
(note the presence of the $\sinh$ function)
\begin{equation}
\d s^2_{4+1} = 
-  F(r)\; \d t^2 
+{\d r^2\over F(r)} 
+ r^2 \; \left[\d\chi^2 + 
\sinh^2\chi \; (\d\theta^2 + \sin^2\theta \; \d\phi^2)\right].
\end{equation}
with (note the presence of the leading {\em minus} $1$!)
\begin{equation}
F(r) =  -1- {2M_{4+1}\over r^2} 
+ {Q_{4+1}^2\over r^4} - {\Lambda_{4+1} \; r^2\over 6}.
\end{equation}
This metric also satisfies the (4+1)-dimensional Einstein--Maxwell
equations, but with a hyperbolic symmetry instead of either
hyper-spherical or hyper-planar symmetry. You can now re-do the
analysis of this note by placing a spatially hyperbolic D-brane
boundary at $r=a(t)$ and will again obtain very similar results to
those of this article.  The intrinsic geometry of the D-brane will now
be
\begin{equation}
\d s^2_{3+1} = - \d \tau^2  + a^2(\tau) \;  \left[\d\chi^2 + 
\sinh^2\chi \; (\d\theta^2 + \sin^2\theta \; \d\phi^2)\right].
\end{equation}

For all three cases, $k=+1,0,-1$, the formal dynamical equation for
the brane motion [equation (\ref{E:formal}), valid for arbitrary
$F(r)$] is unchanged, while the explicit dynamical equation after
{\RNdS} surgery (the generalized Friedman equation) becomes
\begin{equation}
\left({\dot a \over a}\right)^2 = 
- {k\over a^2} 
+ {2M_{4+1}\over a^4} 
-  {Q_{4+1}^2\over a^6} 
+ {\Lambda\over 3}\;
+\left({ 8\pi \; G_{3+1}\over3}\right) 
\left[\rho + \half {\rho^2\over\rho_0} \right]. 
\end{equation}
%

%------------------------------------------------------------------------------
\clearpage
%------------------------------------------------------------------------------
 
%------------------------------------------------------------------------------
\end{document}